\documentclass[a4paper]{ESASPCS13Style}
\usepackage{epsfig}

\begin{document}

\title{Formation and Evolution of Very Low Mass Stars and Brown Dwarfs}

\author{Jochen Eisl{\"o}ffel\inst{1} \and Subhanjoy Mohanty\inst{2}
  \and Alexander Scholz\inst{1}} \institute{Th{\"u}ringer Landessternwarte
  Tautenburg -- Sternwarte 5, D-07778 Tautenburg, Germany,
  \and Harvard-Smithsonian Center for Astrophysics -- 60 Garden Street, Cambridge, 
  MA 02138, USA}

\maketitle 

\begin{abstract}

The formation and evolution of brown dwarfs are currently "hot topics"
in cool star research. Latest observations and modeling efforts on disks, 
accretion, outflows, spatial distribution, and binarity in the context of 
the formation by ejection model and alternative scenarios were presented 
and discussed vigorously in this splinter session. A second major topic was 
rotation and activity of these objects. This part included observations of 
rotation periods and amplitudes, variability, and X-ray activity, and 
their consequences for the generation of magnetic fields and the cool 
atmospheres.

\end{abstract}

\section{Introduction}
  
In recent years, large populations of objects with masses between stars 
and planets have been found in the solar neighbourhood, in young open clusters, 
and in star-forming regions. The formation and evolution of these very low mass
(VLM) stars and brown dwarfs (BDs) is one of the current "hot topics" in cool star 
research. This splinter meeting was intended to bring together the community to 
discuss latest results and future directions of research in this rapidly evolving 
field. 

After the discovery of large numbers of VLM objects one of the most debated 
questions is that of their origin. There are two lines of formation models; scenarios
where brown dwarfs form very similar to stars and models which include ejection from a 
multiple system. Initial observations to distinguish between these proposed scenarios are 
becoming available now. They are summarised and discussed in Sect. \ref{formation}. 

Observations also deliver information on the rotation and activity of these objects. 
These parameters are directly connected to the interior structure and the atmospheres, 
and give insights into the physics of dynamo generation, magnetic fields, surface spots, 
and winds. The talks and discussions of these topics are summarised in Sect. \ref{rotact}.
  
\section{The formation of brown dwarfs}
\label{formation}

A large part of this splinter session was dedicated to an extended discussion
of formation scenarios for brown dwarfs. Today, we distinguish two
main groups of models: BDs could form very similar to stars, i.e. from the direct 
graviational collapse of dense cores (e.g., \cite{padoan02}). Alternatively, 
the formation process could involve a truncation or cut-off of the gas accretion 
reservoir. This cut-off could result either from an ejection from a multiple system
(e.g., \cite{reipurth01}, \cite{bate03}) or from external forces, e.g. the radiation
pressure of a nearby OB star (\cite{whitworth04}). 

The ejection scenarios are of particular interest for observers, because any 
ejection process should significantly influence the properties of the objects. 
Thus, the ejection models can be probed by searching for a signature of the ejection, 
for example by investigating accretion disks, spatial distribution, and binarity 
of young BDs. In the splinter session, all these three subtopics were
discussed in detail.

\subsection{Disks and accretion}

\begin{figure}[t]
  \begin{center}
    \epsfig{file=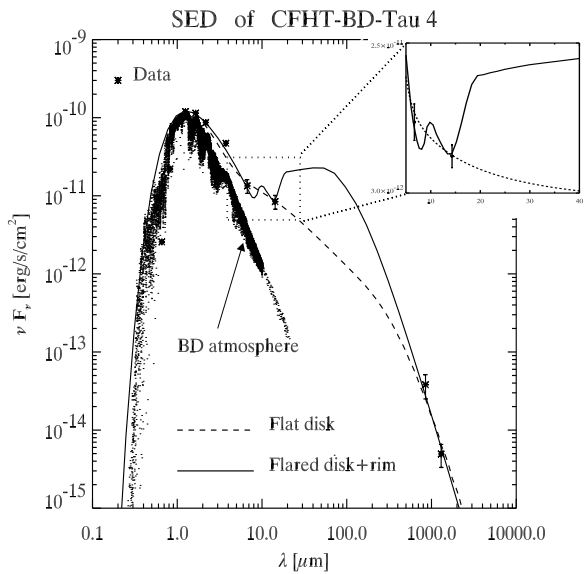,width=8cm}
  \end{center}
\caption{Spectral energy distribution of CFHT-BD-Tau-4 in comparison with model fits
(Pascucci et al. 2003).
The measurements are dereddened with a visual extinction of 3\,mag and are plotted with 
asterisks. The diagram combines data points in the optical, the near- and mid-infrared, 
as well as sub-mm- and mm-values. The atmosphere of a 1\,Myr old BD with 3000\,K effective 
temperature is over-plotted in dots. Apparently, the flux from the BD atmosphere alone 
cannot explain the measurements. The best fitting models are the flat disk (dashed line) 
and the flared disk with an inner puffed-up rim (solid line).\label{apai}}
\end{figure}

The existence of accretion disks around substellar objects has been established
by infrared photometry, optical spectroscopy, and photometric monitoring. In his 
talk, Daniel Apai focused on the results of recent 
infrared observations, which can be used to infer the geometry and structure
of disks. First evidence for circumsubstellar disks came from deep 
mid- and near-infrared surveys in star forming regions and young open clusters
(e.g., \cite{muench01}, \cite{liu03}, \cite{jay03}). For a few objects, the 
combination of NIR and MIR photometry were used to constrain models for the 
disk structure (\cite{natta01}, \cite{apai02}, \cite{mohanty04}). It appears  
that BDs may show a variety of disk geometries, either flared and flat disks, in 
some cases with a central hole. Including sub-mm/mm data points, it was possible 
to constrain the mass for a circumsubstellar disk around CFHT BD Tau 4 to a value 
between 0.4 and 5.7$\,M_\mathrm{Jup}$ (\cite{pascucci03}, see Fig. \ref{apai}). 
For the same object, is has been found that the dust is indeed confined to a disk, 
and that the grains grow to $\approx 2\,\mu m$ sizes (\cite{apai04}). Thus, 
sub-mm/mm observations are probably the best tool to investigate the overall 
geometry of BD disks and to derive disk masses and radii. 

Infrared colour excess is not the only observational signature for the existence
of accretion disks around young BDs. Alexander Scholz presented photometric
light curves of a sample of VLM stars and BDs in young open clusters in Orion,
which show the typical variability characteristic of classical T Tauri stars,
i.e. high amplitude and partly irregular flux modulation. The usual interpretation
for this behaviour is the existence of hot spots formed by matter flow from
the accretion disk onto the object. Most of these highly variable VLM objects
show a near-infrared colour excess. Moreover, optical spectroscopy reveals
a strong accretion signature in their spectra, i.e. intense emission features
like H$\alpha$ and the infrared Ca triplet. Thus, these objects are the
VLM representatives of classical T Tauri stars (\cite{scholz04a}, \cite{scholz04b}). 

Furthermore, Matilde Fern\'andez demonstrated in her talk that VLM objects do also
show outflow activity. Optical spectroscopy for Par-Lup\,3-4, a VLM star near
the substellar limit (\cite{comeron03}), exhibits double-lined [SII] emission 
features, a clear hint for the existence of a jet. Deep imaging revealed that 
there is indeed an [SII] emission feature close to this object, which could
be the first image of an outflow from a VLM object.

All these results demonstrate that brown dwarfs undergo a
T Tauri phase very similar to stars. Particularly, young BDs possess
accretion disks with lifetimes between 3 and 10\,Myr. Although the
disk lifetime of BDs is still very uncertain, it seems to be not vastly 
different from solar-mass stars. However, this result alone is not sufficient
to distinguish between formation scenarios. In particular, it does not
to rule out an ejection scenario, as pointed out by several speakers. 
Ejection should produce truncated disks, but it is not clear, how much 
mass the disk would lose in such a process. Additionally, the processes
which regulate the disk lifetime (i.e. accretion, winds, outflows) are
only poorly understood for VLM objects. 

\subsection{Spatial distribution}
\label{spatial}

If brown dwarfs form by ejection from multiple systems, the ejection
velocities should influence the spatial distribution of these objects.
Therefore, large-scale surveys of star forming regions are able to set
tight constraints for the formation models. 

In this splinter session, two speakers focused on the spatial distribution
of substellar objects in the Chamaeleon star forming region. Jochen Eisl{\"o}ffel
reported about a wide-field survey which delivered more than 100 new VLM members
in Chamaeleon I (\cite{lopez04}). It was shown that this spatial distribution
does not depend significantly on object mass: The young brown dwarfs as well as 
their solar-mass siblings are concentrated in the regions around the two dark
clouds in this region. The survey constrains the average velocity of BDs to 
values below $\sim $1\,km\,s$^{-1}$. Thus, there is no evidence for 
extraordinarily high escape velocities in the substellar regime.

This important finding is confirmed by the results presented by Kevin Luhman.
He pointed out that with ejection velocities up to 10\,km\,s$^{-1}$, the brown dwarfs 
should populate a region with radius 3\,pc around the cloud cores in Chamaeleon.
Large-scale surveys, however, show no extended BD population in this region. 
There is no halo of high-velocity BDs around the cloud cores, in 
contradiction to the original ejection model (\cite{reipurth01}). We conclude that
if BDs form via an ejection process, it should release the BDs at relatively 
low velocities.

\subsection{Binarity}
\label{binarity}

The ejection process should also leave its imprint in the binarity of brown 
dwarfs. As shown by recent HST imaging surveys, the binary frequency among
BDs is in the range of 15\%. Remarkably, there is a paucity of wide BD
binaries with separations larger than 15\,AU compared to solar-type stars
(\cite{martin03}, \cite{bouy03}). Thus, the distribution of separations could be 
either compressed or truncated compared with the distribution for G dwarfs. 
To distinguish between these two possibilities, it is necessary to evaluate 
whether there is a population of short period BD binaries. This was the topic 
of the talk by Ansgar Reiners. He presented results of a search for spectroscopic
binaries among VLM stars and BDs. For about 60 objects, second epoch radial
velocities were measured. With sensitivities of 300\,m\,s$^{-1}$ the minimum 
separations of HST imaging were detectable. Combining radial velocity
data with high-resolution imaging, the complete distribution of 
separations can be covered. Thus far, it appears that short-period binaries
exist among VLM objects, although their frequency is not very high.
This result points in the direction of a cut-off at wide separations, which 
could be interpreted as an indication for an ejection process.

On the other hand, there is evidence for the existence of a few wide 
VLM binary systems in Chamaeleon I star forming region. Within the
wide-field survey presented by Jochen Eisl{\"o}ffel, two possible 
wide binaries with separations around $10''$ were found, which include
objects around and even below the substellar boundary (\cite{lopez04}). 
In his talk, Kevin Luhman now showed images and spectra for a BD binary
in the same region. Colours, spectral types, particularly the low gravity
of both components are consistent with their membership to the very young 
population in this region. Their separation is $1\farcs44$, which corresponds
to 240\,AU. Given the low object density in Chamaeleon, it is unlikely
that the objects are unrelated late-type members of this star forming
region (\cite{luhman04}). 
 
\subsection{Discussion}

Every scenario for the formation of brown dwarfs should be consistent
with the observational results given above. To probe the initial conditions
of the formation process, increasingly sophisticated 3D Smoothed Particle 
Hydrodynamic simulations are now available, as pointed out by J{\"u}rgen
Steinacker in his talk. In combination with 3D continuum radiation transfer,
these SPH simulations can be used to produce images of the collapsing core 
at different wavelengths and evolution times. Thus, the evolution of cloud
cores can be visualised and directly compared with observations (\cite{stein04}).
This method has already turned out to be valuable in the massive star regime. 
Like for brown dwarfs, the question whether accretion through a disk is the main 
formation mechanism is also the key topic in massive star-formation. Recently,
the first massive disk seen in absorption has been detected in M17 (\cite{chini04}). 
The ongoing modeling indicates that the flattened structure around the central 
object may indeed have an extent of about 20000\,AU with a mass around 10 solar 
masses. The same approach is in principle able to probe disk formation 
models in the VLM regime.

In the later stages of the formation process, it is crucial to distinguish
between the star-like formation models and scenarios, where a cut-off from
the gas accretion reservoir is included. The most popular scenario for such a 
cut-off is the ejection process, but there are alternative processes which could
have a similar effect, as explained by Hans Zinnecker in his talk. He explored
the possibility that a pre-stellar core is eroded by the ionising radiation
of a nearby OB star. In principle, such a process could form a brown dwarf
or even planetary mass object from a core, which would otherwise have formed
an intermediate or low-mass star (\cite{whitworth04}). It was concluded that
this photo-erosion scenario works well to produce free-floating VLM objects,
because it works over a wide range of conditions. On the other hand, it is
probably quite inefficient, since it requires a relatively massive initial
core. Moreover, this process can only operate in the vicinity of an OB star,
whereas BDs were found also in regions where such stars are not available.
Thus, the photo-erosion scenario could be an additional process to produce
a small number of BDs, but the predominant part of the substellar population
must have a different origin.

Therefore, the pivotal point for our understanding of BD formation is
still to distinguish between scenarios with or without an ejection process.
This problem was intensely debated during the general discussion in the splinter 
session. The observational results discussed above delivered valuable
guidance through this discussion. One main prediction of the original
ejection model are high spatial velocities, which would produce a spherical
halo of BDs in star forming regions (\cite{reipurth01}). Wide-field surveys 
do not see such a halo, as pointed out in Sect. \ref{spatial}. However, recent 
simulations show that an ejection process does not necessarily lead to very high 
spatial velocities (\cite{umbreit04}), which makes it difficult to rule out this 
scenario based on the spatial distribution of BDs. 

It is also very complicated to constrain the formation scenarios by investigating
the accretion disks, since the ejection scenario does not rule out the existence
of long-lived disks. Thus, it could be that we just observe the BDs, whose disks
have survived the ejection. Our best hope in this field are sub-mm observations,
which could deliver disk masses for large samples of BDs. 

Finally, binarity surveys are a third important tool to verify BD formation scenarios.
Two conflicting observational results have been presented in this splinter 
session. On the one hand, several groups find a cut-off of the BD binary distribution
at separations of about 15\,AU (e.g., \cite{martin03}, \cite{bouy03}). It is not clear
whether this finding is just an observational bias or an indication for the truncation
of wide binaries by ejection. On the other hand, there are a few examples of wide
binaries detected in star forming regions, as pointed out in Sect. \ref{binarity}.
Future astrometric space missions are probably able to establish beyond doubt if
these objects are indeed physically linked. But since there are only 
very few of these wide binary BDs, it is doubtful whether their existence tells us 
something about BD formation in general. At least, these objects might indicate that 
BD formation is possible without an ejection process.

For the future research in this field, the investigation of the earliest stages of
the formation process will become more and more important. The detection of 
'proto' brown dwarfs could indeed deliver clear evidence for a star-like
origin of these objects. Subhanjoy Mohanty pointed out in the discussion that very 
recent (as-yet unpublished) radio observations appear to indicate the presence of at 
least some gravitationally bound cores with masses in the substellar regime (Walsh 
et al., in prep.). The confirmation of these results would support star-like formation 
for brown dwarfs.

Recapitulating, it is still too early for a definite answer about the process that
forms substellar objects. Young brown dwarfs share many properties with their 
more massive siblings, in particular disk frequency, disk geometry, and spatial
distribution, indicating that the majority of these objects forms in a way similar
to stars. On the other hand, there is no clear observational result which can only
be explained by including an ejection process. Thus, many of the participants suggested
that the ejection scenarios are not of significant importance for BD formation. 

\section{Rotation and activity of VLM objects}
\label{rotact}

The second main focus of the splinter session was the rotation
and activity of VLM objects. This subject area was covered by four talks,
which will be reviewed in the following. 

Subhanjoy Mohanty presented results of a high resolution spectroscopic
study of young brown dwarfs. Particularly, he concentrated on the
interdependence of rotational velocity and H$\alpha$ 10\%
width, where the latter is directly correlated with the accretion rate. 
A plot of $v\sin{i}$ vs. H$\alpha$ shows a conspicuous morphology: Whereas
slow rotators cover a broad range of H$\alpha$ widths up to 400\,km\,s$^{-1}$,
fast rotators show only very low H$\alpha$ activity, with 10\% widths
below 200\,km\,s$^{-1}$. The boundary between fast and slow rotators
lies at 20\,km\,s$^{-1}$. In other words, while non-accretors (H$\alpha$ 
width $<$ 200\,km\,s$^{-1}$) are found at both high and low velocities, 
accretors (H$\alpha$ width $>$ 200\,km\,s$^{-1}$) appear to be exclusively 
slow rotators. The most straightforward interpretation is
that the accretion disk is able to brake the rotation. It was pointed
out in the discussion that the same result was obtained for VLM objects
in the $\sigma$\,Ori cluster, where a plot of angular velocity inferred from
the rotation period vs. photometric amplitude (which can be used as
accretion indicator) shows a very similar morphology (\cite{scholz04a}). 
Thus, there is evidence for rotational braking through interaction with
the circumsubstellar disk. In addition, the mean rotational velocity 
increases as the objects get older. This can be explained by spin-up 
due to hydrostatic contraction.  

\begin{figure}[t]
  \begin{center}
    \epsfig{file=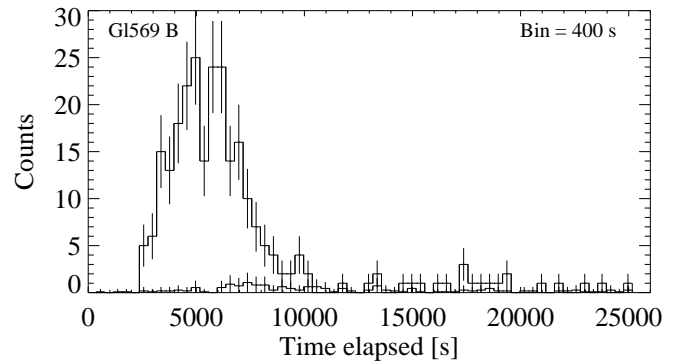,width=9cm}
  \end{center}
\caption{X-ray light curve of the field BD GL569Bab (from Stelzer et al.). 
Apart from a strong flare, the object does probably also show quiescent 
X-ray emission. The lowest curve, which is close to zero, indicates the background. 
If confirmed, this observations would be the first detection of quiescent X-ray 
emission from an evolved BD.
\label{stelzer}}
\end{figure}

In the following talk, Reinhard Mundt reported about an extended
variability and rotational study of young stars in NGC2264 (age 1-3\,Myr), 
which delivered a large sample of about 400 rotation periods (\cite{lamm04}). 
Since the object masses reach down to the substellar limit, this dataset 
is particularly well-suited to probe the mass dependence of rotation
and activity. One main result of this study is a change in the photometric
amplitudes in the VLM regime. At a R-I colour of 1.5, the amplitudes
decrease by a factor of about three. Thus, VLM objects show only very 
low amplitude variations. The most convincing explanation so far is that 
these objects exhibit smaller spots than more massive stars, probably 
because the coupling between magnetic field and gas is significantly 
decreased on these very cool objects. Indeed, the magnetic Reynolds number, 
which should exceed unity to ensure coupling between field and gas, 
goes down from six at spectral type M2 to about 0.8 at M5.5. This is
just the same spectral range in which the photometric amplitudes 
decrease. An alternative interpretation for the low amplitudes would 
be highly symmetric spot distributions, but this would require a change
in the magnetic field configuration, which is unlikely, since at the
age of NGC2264 all stars with masses up to $1\,M_{\odot}$ are fully
convective. 

To probe the magnetic field properties of VLM objects, it is highly
interesting to study their X-ray properties. This was the topic of 
Beate Stelzer's talk. VLM objects are fully convective throughout
their evolution. Thus, a solar-type dynamo cannot work in these objects,
since it requires a radiative core. However, VLM objects at least down to
late M spectral types show X-ray activity: Several BDs in Chamaeleon I have 
been detected by ROSAT (\cite{neu98}). Recent observations with XMM confirmed
these detections, and proved that the ROSAT sources are not spurious
(\cite{stelzer04}). In addition to these Chamaeleon objects, X-ray emission 
was detected for BDs in the Pleiades (\cite{stelzer03}, \cite{briggs04}) and 
in the field (\cite{rutledge00}, \cite{stelzer03}). The X-ray light curve
of the field BD GL569Bab shows that -- apart from flares -- this objects does 
probably also show quiescent emission (see Fig. \ref{stelzer}).
Thus, the dynamo of BDs does persist at least up to
ages of 1\,Gyr. However, no L-dwarf has been detected in X-rays so far.
This seems to be in agreement with the decline of the H$\alpha$ activity
at late M spectral types, which is interpreted as a consequence of the
cool effective temperatures, which result in high atmospheric resistivities
(\cite{mohanty02}). In combination with the results presented by Reinhard 
Mundt, we conclude that VLM objects show magnetic activity for as long as they 
are not too cool to prevent coupling between field and gas.

In the same temperature range, where the magnetic activity declines, another
process begins to play a significant role: the condensation of dust clouds
in the atmospheres. Atmosphere models suggest that this process becomes
important for objects with $T<2500$\,K, corresponding to late M spectral types.
Similar to magnetic spots on solar-like stars, dust clouds could produce
inhomogeneous surfaces, which could cause periodic modulations of the
flux. This was the motivation for numerous variability studies on ultra-cool
dwarfs, which were reviewed in the talk of Coryn Bailer-Jones. Variability
with small amplitudes has been detected on about 40\% of the targets
(e.g., \cite{bailer01}, \cite{clarke02}, \cite{enoch03}). The origin of these 
variations is most likely the existence of dust clouds, since these objects are 
too cool to produce magnetic spots (\cite{gelino02}, see above). Spectrophotometric
monitoring has been used to distinguish between clouds and spots, and it 
delivered tentative evidence for the existence of dust (\cite{bailer02}).
It was surprising that ultra-cool dwarfs in most cases show only non-periodic
variability and no photometric period, which is in contrast to similar studies
for solar-like stars. One explanation could be that the variability surveys
are not sensitive to the periods of ultra-cool dwarfs. However, rotational
velocities show that these objects are in general rapid rotators with
$v\sin{i}=10\ldots 40$\,km\,s$^{-1}$, corresponding to a period range where the 
variability studies were sensitive (\cite{bailer04}). Therefore,
to explain the absence of photometric periods, Bailer-Jones et al.
proposed the 'mask hypothesis': The clouds evolve on timescales of a few hours, 
and thus the period is 'masked' by non-periodic variability (\cite{bailer01}, 
\cite{gelino02}). In the discussion, it was pointed out that there is
indeed a change of the variability characteristic at late M spectral types,
in the sense that earlier type objects often show stable periods. The 
most probable explanation is a change of the nature of the surface features
from magnetically induced spots to dust clouds at late M spectral types.

\begin{acknowledgements}

  The conveners of this splinter session would like to thank all
  speakers and participants for their contributions and the lively
  discussion. It is a pleasure to thank the organisers of the 
  Cool Stars 13 conference, who made this splinter session possible.
  The work of J.E. and A.S. was partially funded by Deutsche 
  Forschungsgemeinschaft (DFG), grants Ei\,409/11-1 and Ei\,409/ 11-2. 
  A.S. acknowledges financial support for travel granted by the European 
  Space Agency (ESA). S.M. would like to thank the SIM-YSO group for funding
  his research.

\end{acknowledgements}

\end{document}